\documentclass[journal=jpclcd,manuscript=article]{achemso}

\usepackage{chemformula} % Formula subscripts using \ch{}
\usepackage[T1]{fontenc} % Use modern font encodings

\author{Mouhui Yan}
\affiliation[SHU]
{Department of Physics, Shanghai University, 200444 Shanghai, China}
\author{Yichen Jin}
\affiliation[SHU]
{Department of Physics, Shanghai University, 200444 Shanghai, China}
\author{Zhicheng Wu}
\affiliation[SHU]
{Department of Physics, Shanghai University, 200444 Shanghai, China}
\author{Arshak Tsaturyan}
\affiliation[SFedU]
{Institute of Physical and Organic Chemistry, Southern Federal University,\\344090 Rostov on Don, Russia}
\author{Anna Makarova}
\affiliation[FUB]
{Institut f\"ur Chemie und Biochemie, Freie Universit\"at Berlin, Arnimallee 22,\\14195 Berlin, Germany}
\author{Dmitry Smirnov}
\affiliation[TUDD]
{Institut f\"ur Festk\"orper- und Materialphysik, Technische Universit\"at Dresden,\\01069 Dresden, Germany}
\author{Elena Voloshina}
\affiliation[SHU]
{Department of Physics, Shanghai University, 200444 Shanghai, China}
\email{elena.voloshina@icloud.com}
\alsoaffiliation[FUB]
{Institut f\"ur Chemie und Biochemie, Freie Universit\"at Berlin, Arnimallee 22,\\14195 Berlin, Germany}
\author{Yuriy Dedkov}
\affiliation[SHU]
{Department of Physics, Shanghai University, 200444 Shanghai, China}
\email{yuriy.dedkov@icloud.com}
\alsoaffiliation[FUB]
{Institut f\"ur Chemie und Biochemie, Freie Universit\"at Berlin, Arnimallee 22,\\14195 Berlin, Germany}

\title[]
{Correlations in the Electronic Structure of van der Waals NiPS$_3$ Crystals:
An X-Ray Absorption and Resonant Photoelectron Spectroscopy Study}

\begin{document}
%%%%%%%%%%%%%%%%%%%%%%%%%%%%%%%%%%%%%%%%%%%%%%%%%%%%%%%%%%%%%%%%%%%%%
%% The "tocentry" environment can be used to create an entry for the
%% graphical table of contents. It is given here as some journals
%% require that it is printed as part of the abstract page. It will
%% be automatically moved as appropriate.
%%%%%%%%%%%%%%%%%%%%%%%%%%%%%%%%%%%%%%%%%%%%%%%%%%%%%%%%%%%%%%%%%%%%%

\begin{tocentry}

\includegraphics[width=\textwidth]{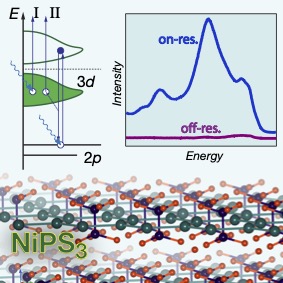}
%Some journals require a graphical entry for the Table of Contents. This should be laid out ``print ready'' so that the sizing of the text is correct.

\end{tocentry}

%%%%%%%%%%%%%%%%%%%%%%%%%%%%%%%%%%%%%%%%%%%%%%%%%%%%%%%%%%%%%%%%%%%%%
%% The abstract environment will automatically gobble the contents
%% if an abstract is not used by the target journal.
%%%%%%%%%%%%%%%%%%%%%%%%%%%%%%%%%%%%%%%%%%%%%%%%%%%%%%%%%%%%%%%%%%%%%
\begin{abstract}

The electronic structure of high-quality van der Waals NiPS$_3$ crystals was studied using near-edge x-ray absorption spectroscopy (NEXAFS) and resonant photoelectron spectroscopy (ResPES) in combination with density functional theory (DFT) approach. The experimental spectroscopic methods, being element specific, allow to discriminate between atomic contributions in the valence and conduction band density of states and give direct comparison with the results of DFT calculations. Analysis of the NEXAFS and ResPES data allows to identify the NiPS$_3$ material as a charge-transfer insulator. Obtained spectroscopic and theoretical data are very important for the consideration of possible correlated-electron phenomena in such transition-metal layered materials, where the interplay between different degrees of freedom for electrons defines their electronic properties, allowing to understand their optical and transport properties and to propose further possible applications in electronics, spintronics and catalysis.

\end{abstract}

%%%%%%%%%%%%%%%%%%%%%%%%%%%%%%%%%%%%%%%%%%%%%%%%%%%%%%%%%%%%%%%%%%%%%
%% Start the main part of the manuscript here.
%%%%%%%%%%%%%%%%%%%%%%%%%%%%%%%%%%%%%%%%%%%%%%%%%%%%%%%%%%%%%%%%%%%%%

\newpage

Layered materials, like graphene, silicene or transition metal dichalcogenides (TMDs) and different heterosystems on their basis have attracted much attention in the last years~\cite{Geim:2009,DasSarma:2011br,Geim:2014hf,Zhuang:2015cz,Carvalho:2016abc,Manzeli:2017abc}. These studies are motivated by the hope on the possible new applications of these materials in different areas, where their unique electronic and transport properties might bring new functionalities. These layered materials can be relatively easily produced in the high quality form either as bulk materials or as a monolayer (or multilayer) on different substrates and then isolated  for further studies or applications~\cite{Bae:2010,Chhowalla:2013fz,Ryu:2014fo,Dedkov:2015kp,Li:2018fz,Dedkov:2020da}. Such high-quality layers and heterosystems made from layered 2D materials allow to perform studies of different intriguing collective quantum phenomena, which appear in low-dimensional systems, like, for example, superconductivity for twisted graphene bi-layers~\cite{Cao:2018ff,Andrei:2020abc}, charge density waves in TMDs~\cite{Xi:2015jz,Hossain:2017fr}, and coherent excitonic excitations in antiferromagnetic transition metal trichalcogenide (TMT) NiPS$_3$~\cite{Kang:2020abc}. For example, recently, different exciting phenomena in 2D materials were extended to the observation of the long-range Ising-type antiferromagnetic (AFM) order in the TMT FePS$_3$ monolayer with a N\'eel temperature ($T_N$) of $118$\,K~\cite{Lee:2016ga,Wang:2016ez}. Following these works, a layer-dependent ferromagnetic (FM) order in CrI$_3$ with a Curie temperature ($T_C$) of $45$\,K was reported~\cite{Huang:2017kd} and then the FM magnetic order has been observed for single atomic layers of Cr$_2$Ge$_2$Te$_6$ ($T_C\sim40$\,K)~\cite{Gong:2017jf}, Fe$_3$GeTe$_2$ ($T_C\sim220$\,K)~\cite{Deng:2018cp}, and MnSe$_2$~\cite{OHara:2018hi}. These inspiring discoveries of magnetic order in 2D monolayers open up exciting opportunities for the application of these properties in new transport and optical devices and for the exploration of intrinsic 2D Ising-type (anti)ferromagnetism.

\begin{figure}
\center
\includegraphics[width=0.65\columnwidth]{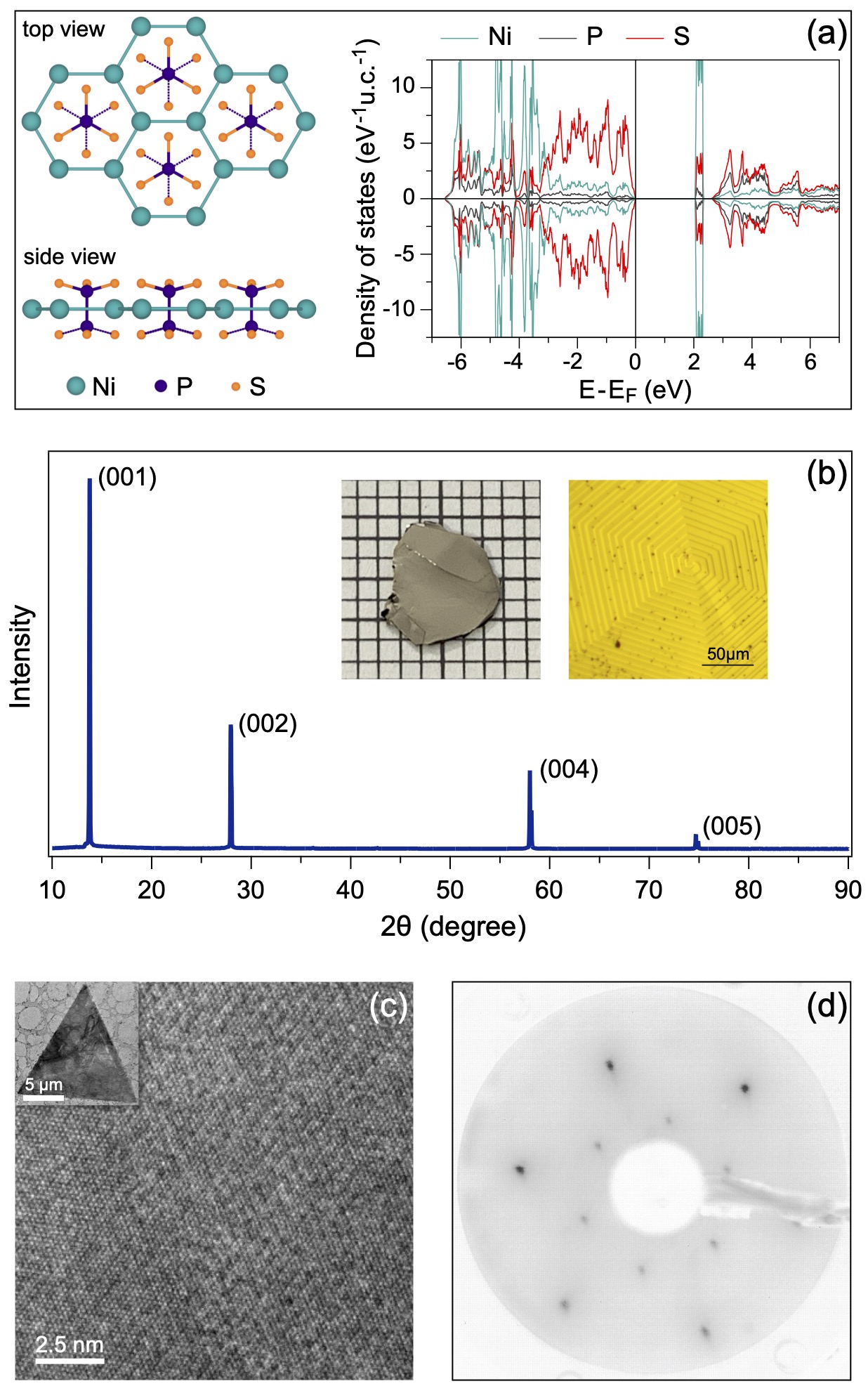}\\
\vspace{1cm}
\caption{(a) Top and side views of the crystallographic structure of the NiPS$_3$ single layer. Atom-projected partial DOS for bulk NiPS$_3$ in the AFM ground state is shown on the left-hand side of the panel. (b) XRD patterns of NiPS$_3$ measured at room temperature. Insets show photo of the sample and its optical microscopy image, respectively. (c) TEM image of the NiPS$_3$. Inset shows the respective large scale TEM image of several-$\mu\mathrm{m}^2$ NiPS$_3$ sheet. (d) MCP-LEED image of the NiPS$_3$(001) surface acquired at $95$\,eV.}
\label{fig:structure}
\end{figure}

The above mentioned layered transition metal trichalcogenides (TMPX$_3$; TM: $3d$ TMs, like Fe, Co, or Ni; X: chalcogen, like S or Se) came to the focus of the studies of 2D materials only recently~\cite{Susner:2017in,Wang:2018dha,Zhu:2020db}. From the crystallographic point of view they can be seen as layers of TMS$_2$ where one third of TM atoms is replaced by P--P dimers perpendicular to the layer (Fig.~\ref{fig:structure}(a)) and in the 3D structure they crystalize into the $C2/m$ or $R\bar{3}$ space groups~\cite{Du:2016ft,Yang:2020ex}. Most of these compounds are AFM wide-gap semiconductors~\cite{Sivadas:2015gq,Chittari:2016cda,Yang:2020ex} and they were proposed for different applications like gas sensors, optical sensors, effective materials in water splitting applications and in (opto)spintronics~\cite{Li:2014de,Zhang:2016kra,Zhu:2020db}. Surprisingly, at the present time, the main focus in the studies of this class of materials is concentrated on the applied-oriented investigations, missing the systematic studies of their electronic structure using different electron spectroscopy methods, which provide direct insight in these properties. Such experiments can give a direct comparison with the available theoretical data and link them to the available application-oriented studies, providing the route to improve their properties~\cite{Yang:2020ko}.

Here, we present systematic x-ray photoelectron spectroscopy studies of the representative TMT layered material, NiPS$_3$. Using density functional theory (DFT) calculations, it is shown that NiPS$_3$ is an AFM semiconductor in its ground state. It has a band gap of more than $2$\,eV and the magnetic properties of this material are described with a model Hamiltonian of the XY-type. Near-edge x-ray absorption spectroscopy (NEXAFS) and resonant photoelectron spectroscopy (ResPES) performed around the respective absorption edges allow to discriminate between different atomic contributions in the density of states (DOS) for valence and conduction bands. Our analysis of the NEXAFS and ResPES data for the Ni $L_{2,3}$ absorption edge indicate that NiPS$_3$ can be described as a charge-transfer insulator according to the Zaanen-Sawatzky-Allen scheme with strong correlation effects in the valence band of this material. Details describing experimental and theoretical approaches used in the present study as well as additional data are presented in the Supplementary Information.

Layered NiPS$_3$ bulk crystals in our study were synthesized using chemical vapor transport (CVT) method from the stoichiometric amounts of elements and obtained several-$\mathrm{mm}^2$ samples were characterized using different bulk- and surface-sensitive techniques. Fig.~\ref{fig:structure}(b,c) show x-ray diffraction (XRD) plot, optical images and transmission electron microscopy (TEM) data for NiPS$_3$ crystal and these data are in perfect agreement with referenced and previously published data~\cite{Du:2016ft,Jenjeti:2018fu,Xue:2018eq} demonstrating high-quality of samples studied in the present work (see also Fig.\,S1 and  Fig.\,S2 in Supplementary Information for the high-resolution cross-section TEM images and energy dispersive x-ray (EDX) analysis data collected in the TEM measurements, respectively). The distance between single layers of NiPS$_3$ extracted from the high-resolution TEM data is $6.357$\,\AA, which is in very good agreement with the value of $6.301$\,\AA\ obtained in our DFT calculations for fully relaxed 3D structure. [The complete crystallographic data for the fully optimized 3D structure of NiPS$_3$ are presented in Table\,S3.] Raman spectra of NiPS$_3$ are shown in Fig.\,S4 of Supplementary Information and also in very good agreement with previously presented results, which were used for the assignment of the observed scattering peaks~\cite{Bernasconi:1988dj,Kuo:2016ksa,Jenjeti:2018fu,Kim:2019gj}.

NiPS$_3$ was characterized using different surface-sensitive methods and Fig.~\ref{fig:structure}(d) presents low-energy electron diffraction (LEED) image of the freshly-cleaved and UHV-degassed sample. It shows clear hexagonal diffraction spots, which are due to the hexagonal unit cell of NiPS$_3$ in the $xy$-plane perpendicular to the impinge electron beam. The presented LEED image was collected in the low-current electron-beam regime at the primary electron energy of $95$\,eV indicating high surface crystallographic quality with long-range order. Our reference measurements using Au-crystal allow to conclude that presented image shows 1st and 2nd diffraction orders for the NiPS$_3$ $xy$-surface.

\begin{figure}
\center
\includegraphics[width=\columnwidth]{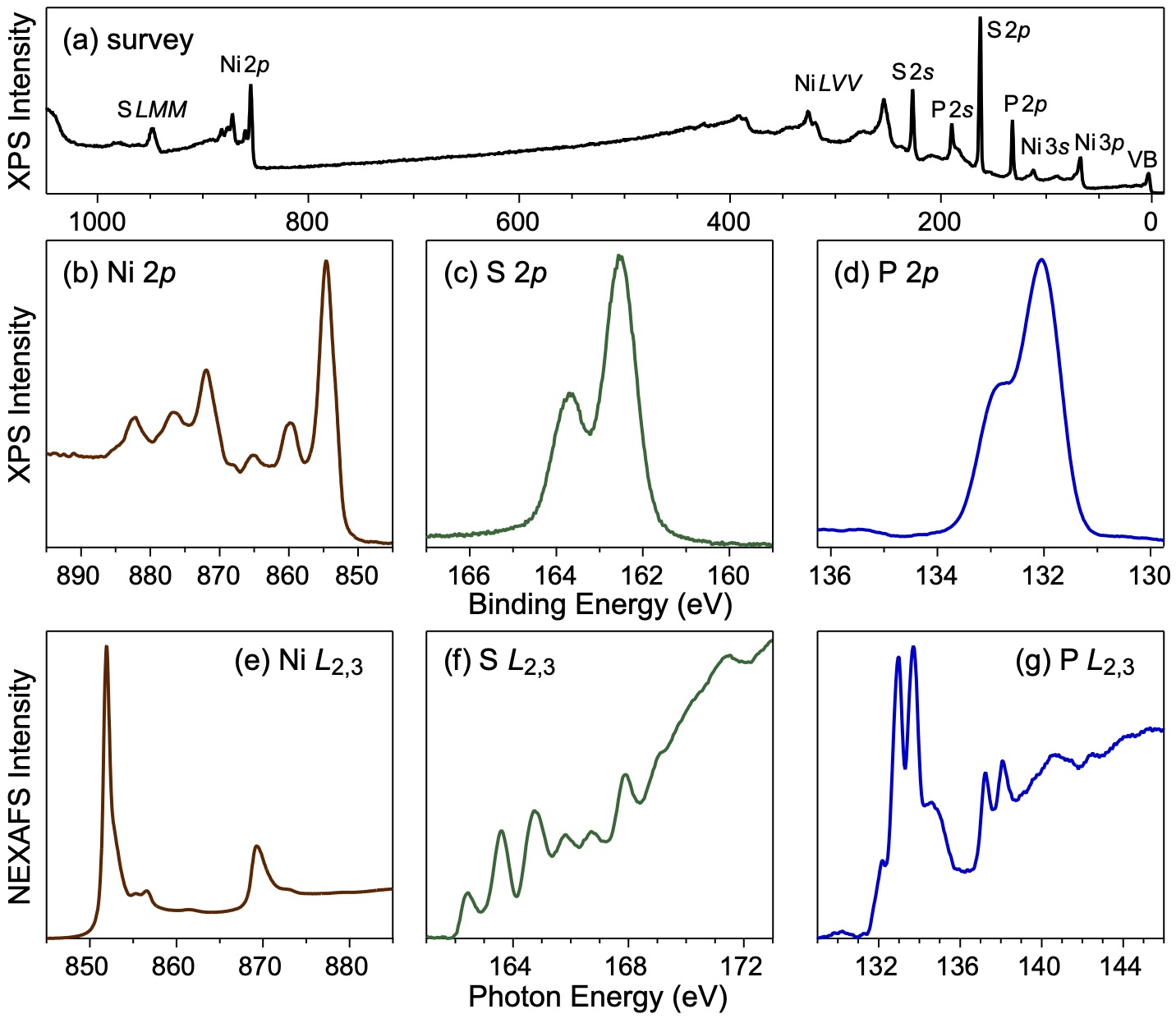}\\
\vspace{1cm}
\caption{XPS spectra of NiPS$_3$ collected at $h\nu=1100$\,eV: (a) survey, (b) Ni\,$2p$, (c) S\,$2p$, (d) P\,$2p$. NEXAFS spectra of NiPS$_3$ collected in the TEY mode: (e) Ni\,$L_{2,3}$, (f) S\,$L_{2,3}$, (g) P\,$L_{2,3}$.}
\label{fig:XPS_NEXAFS}
\end{figure}

The results of x-ray photoelectron spectroscopy (XPS) characterization and NEXAFS measurements at the Ni, S, and P $L_{2,3}$ absorption edges for NiPS$_3$ are summarized in Fig.~\ref{fig:XPS_NEXAFS}. Survey XPS spectrum measured for NiPS$_3$ shows a series of characteristic emission peaks (without any additional emission lines from possible contaminants), which can be easily identified, and they are marked in panel (a). The characteristic multiplet structure for the Ni\,$2p$ spectrum (Fig.~\ref{fig:XPS_NEXAFS}(b)) is in very good agreement with the recent data for NiPS$_3$~\cite{Kim:2018gj} and NiGa$_2$S$_4$~\cite{Takubo:2007kc}, where multiplet modeling using the NiS$_6$ cluster gives the dominant Ni\,$d^9\underline{L}$ ground state in both cases ($\underline{L}$ marks the hole on the ligand site). S\,$2p$ and P\,$2p$ XPS spectra (Fig.~\ref{fig:XPS_NEXAFS}(c,d)) demonstrate clear single spin-orbit-slit doublets indicating strongly localized covalent bonds between P and S atoms forming (P$_2$S$_6$)$^{4-}$ anions.

\begin{figure}
\center
\includegraphics[width=0.6\columnwidth]{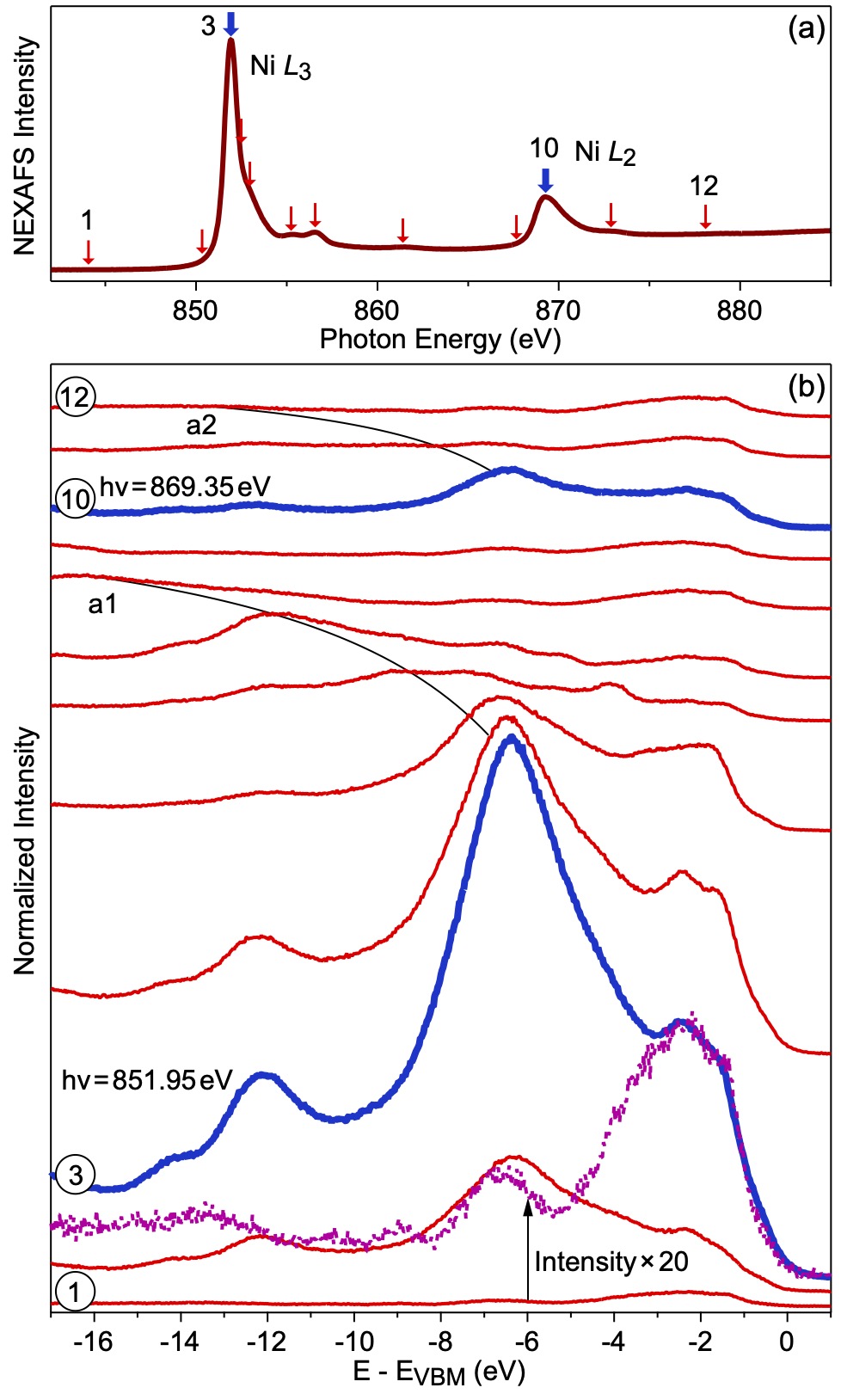}\\
\vspace{1cm}
\caption{Results of ResPES experiments for NiPS$_3$: (a) the reference Ni\,$L_{2,3}$ NEXAFS spectrum and (b) a series of photoemission spectra taken at the particular photon energies marked by the corresponding vertical arrows in panel (a). Spectrum measured at the off-resonance photon energy (spectrum 1) with intensity multiplied by factor 20 is also shown by the doted line. Solid thin curves drawn on top of the spectra $3-12$ and marked by a1 and a2 note the respective Auger-decay lines. All ResPES spectra are shifted in vertical direction for clarity. Zero of the energy scale for ResPES data is referenced to the valence band maximum edge for NiPS$_3$ extracted from the off-resonance spectra.}
\label{fig:ResPES}
\end{figure}

The Ni\,$L_{2,3}$ NEXAFS spectrum (Fig.~\ref{fig:XPS_NEXAFS}(e) and Fig.~\ref{fig:ResPES}(a)) demonstrates characteristic absorption edge for two spin-orbit split components separated by $\approx 17$\,eV. According to the Zaanen-Sawatzky-Allen scheme for transition-metal compounds (insulators and semiconductors)~\cite{Zaanen:1985bs}, their electronic structure can be described with two parameters, the $d-d$ correlation energy ($U_{dd}$) and charge transfer energy between $d$-states of metal and $p$-states of ligand ($\Delta$). Simplifying all considerations, one can say that if $U_{dd}<\Delta$, then compound is the Mott-Hubbard type insulator, while it is the charge-transfer type insulator if $U_{dd}>\Delta$. For NiPS$_3$ the Ni\,$L_{2,3}$ NEXAFS spectra were simulated in Ref.~\citenum{Kim:2018gj} and good agreement with experimental data was suggested for $U_{dd}=5$\,eV and $\Delta=-1$\,eV, indicating that this compound is the so-called self-doped negative charge-transfer insulator. In our simulations (see computational details and Fig.\,S5 in Supplementary Information) using $U_{dd}=6$\,eV, the best fit is achieved for charge transfer energy between $\Delta=0.5$\,eV and $\Delta=1$\,eV giving simple charge-transfer insulating state for NiPS$_3$. Besides the correctly shown two-peaks satellite structure above the $L_3$ absorption edge, with these parameters our simulations also correctly reproduce the high-energy shoulder for these line as well as all fine structures in the NEXAFS spectra. Our DFT calculations with $U_{dd}=6$\,eV lead to the correct crystallographic structure of NiPS$_3$ with simultaneous adequately reproduced value for $T_N=148$\,K (vs. $T_N=155$\,K obtained in the experiment~\cite{LeFlem:1982gs}). In this case, the respective value for the occupation number for $d$ orbitals of the Ni$^{2+}$ ion is $8.2$ and the calculated value $1.6\,\mu_B$ for magnetic moment of Ni ions is closer to the formal value of $2\,\mu_B$ compared to the previously reported results~\cite{Kim:2018gj}. Our DFT calculations also give significantly wider band gap for NiPS$_3$ -- 2.19\,eV (vs. $\approx1$\,eV in Ref.~\citenum{Kim:2018gj}) together with correctly calculated optical absorption spectra (see Fig.\,S6 in Supplementary Information). Also our DFT calculations give the different distribution of the PDOS in the valence band compared to the previous results (see Fig.~\ref{fig:structure}(a) and Fig.\,S7 for present DFT results obtained using PBE$+U+$D2 and HSE06+D2 functionals, respectively). One has to note that value $\Delta=0.95$\,eV was recently obtained during correct simulation of the resonant inelastic x-ray scattering spectra of NiPS$_3$~\cite{Kang:2020abc} supporting our results.

The S\,$L_{2,3}$ and P\,$L_{2,3}$ NEXAFS spectra of NiPS$_3$ shown in Fig.~\ref{fig:XPS_NEXAFS}(f) and (g), respectively, have a very rich structure consisting of sharp peaks. These spectra reflect the transitions from the ligand spin-orbit split $2p_{3/2}$ and $2p_{1/2}$ core levels onto the unoccupied $d$ and $s$ states. The interpretation of these spectra is quite tedious and ambiguous procedure, however one can conclude that the sharp double-peak structure in the S\,$L_{2,3}$ ($163.6$\,eV and $164.75$\,eV) and P\,$L_{2,3}$ ($132.95$\,eV and $133.7$\,eV) spectra reflects the electron transitions into the first unoccupied hybrid $3s$-like antibonding state formed by S and P~\cite{Farrell:2002ei,Kruse:2009ed}. The first peak in the S\,$L_{2,3}$ spectrum at $162.45$\,eV and the respective shoulder in the P\,$L_{2,3}$ spectrum at $132.2$\,eV most probably correspond to the states in the conduction band peaked at $E-E_{F}\approx2.4$\,eV. Further peaks in the NEXAFS spectra observed at higher photon energies can be assigned either to the so-called ``echo'' or shadow effect of the spin-orbit split features due to multiple scattering or to electron transitions to a mixed-valence band states~\cite{Farrell:2002ei,Kruse:2009ed,Yang:2012iq}.

The results of the ResPES measurements at the Ni\,$L_{2,3}$ absorption edge for NiPS$_3$ are presented in Fig.~\ref{fig:ResPES}, supporting the description of this material as the charge-transfer insulator: In panel (a) the reference Ni\,$L_{2,3}$ NEXAFS spectrum is shown and in panel (b) a series of photoemission spectra taken at the particular photon energies marked by the vertical arrows in (a) is presented. Previous considerations of the transition metal oxides and sulfides~\cite{Fujimori:1984ju,Fujimori:1990gk,Okada:1992cv,Tjernberg:1996hz,Taguchi:2008jj} show that in case of the Mott-Hubbard insulator with $U_{dd}<\Delta$ (e.\,g., Mn and Fe oxides) the core-level and valence band XPS spectra consist of the low-binding-energy intense peak originated mainly from the $3d^{n-1}$ final state and the satellite $3d^n\underline{L}$ structure at larger binding energies. For the charge-transfer insulator with $U_{dd}>\Delta$ (e.\,g., Ni oxide), the low-binding-energy intense peak is formed by the configuration mixing of the $3d^n\underline{L}$ and $3d^n\underline{Z}$ final states ($\underline{Z}$ is the bound states; e.\,g., for NiO its a Zhang-Rice doublet bound states)~\cite{Taguchi:2008jj} and the peak associated with the $3d^{n-1}$ final state is located at larger binding energies with smaller intensities. (Here one has to note that strong hybridization between two final states does not allow to clearly separate them in the photoemission spectra of the valence band region, requiring detailed fitting of the experimental data using cluster model calculations.)

In the ResPES measurements at the Ni\,$L_{2,3}$ absorption edge for NiPS$_3$ the photoemission intensity is the result of interference of two emission channels~\cite{Davis:1981kw}: (i) a direct photoemission $2p^63d^n+h\nu\rightarrow2p^63d^{n-1}+e$ and (ii) a photoabsorption process followed by a participator Coster-Kronig decay $2p^63d^n+h\nu\rightarrow2p^53d^{n+1}\rightarrow2p^63d^{n-1}+e$ and the initial and final states for these two photoemission channels are identical. Also the interference between these excitation channels leads to the Fano-type resonance for states with the $3d^{n-1}$ final-state character~\cite{Davis:1981kw}. For the $2p\rightarrow3d$ ResPES experiments the second emission channel (Coster-Kronig decay) prevails in the spectra, therefore giving the position of the $3d^{n-1}$ peak relative to the position of the $3d^n\underline{L}$ line.

The off-resonance spectrum of NiPS$_3$ measured at $h\nu=843.34$\,eV (spectrum $1$ in Fig.~\ref{fig:ResPES}(b) and its intensity multiplied by factor 20 is shown by the dotted line) demonstrates a broad two-peaks structure in the range of $E-E_{VBM}=-5...0$\,eV and additional satellite structure at larger binding energies. If we assume that NiPS$_3$ is a charge-transfer insulator with $U_{dd}>\Delta$ (according to the previous considerations and considering the similar ion state for charge-transfer insulating state of NiO), then the first structure can be assigned to the $3d^n\underline{L}$ final state and other peaks to the $3d^{n-1}$ final state. The on-resonance photoemission spectrum measured with the photon energy $h\nu=851.94$\,eV corresponding to the $L_3$ absorption edge (spectrum $3$ in Fig.~\ref{fig:ResPES}(b)) demonstrates the drastic increase of the photoemission intensity by factor of $20$ for the two-peaks structure in the range $E-E_{VBM}=-5...0$\,eV and by factor of $80$ for the peaks at large binding energies. However, as it was discussed earlier, in the ResPES experiments the $3d^{n-1}$ Coster-Kronig decay channel mainly contributes in the photoemission spectra. Thus we can conclude that the huge peak at $E-E_{VBM}\approx-6$\,eV and the additional satellites structure at $ E-E_{VBM}\approx-12$\,eV correspond to the $3d^{n-1}$ final state and confirm that NiPS$_3$ has to be described as the charge-transfer insulator material. The similar consideration is valid for the on-resonance photoemission spectrum measured at the $L_2$ absorption edge (spectrum $10$ in Fig.~\ref{fig:ResPES}(b)). The obtained results which allow to describe NiPS$_3$ as charge-transfer insulator are supported by the DFT results. In the DOS plot (Fig.~\ref{fig:structure}(a)), the top of the valence band is dominated by the S\,$p$ states with the Ni\,$3d$ states spreading between $-3.5$\,eV and $-6.5$\,eV. At the same time the bottom of the conduction band is formed mainly by unoccupied Ni\,$3d$ states, leading to the description of NiPS$_3$ as a charge-transfer insulator according to the Zaanen-Sawatzky-Allen scheme (see Fig.\,S8) and confirming experimental spectroscopic results. [Here we have to note that the correct simulation of the ResPES spectra on the basis of the available DOS data is only possible in the many-electron approximation, which might be a topic for further studies.] The presented results are in a rather good agreement with the experimental and calculated ResPES spectra of NiO (Ni$^{2+}$) in the charge-transfer insulating state, where similar spectral behavior was found~\cite{Tjernberg:1996hz,Taguchi:2008jj}.

In summary, we performed systematic studies of the crystallographic structure and electronic properties of layered van der Waals material NiPS$_3$ using different \textit{ex situ} and \textit{in situ} UHV surface science techniques, including LEED and electron spectroscopy methods, like XPS, NEXAFS and ResPES. Our structural investigations demonstrate very high quality of the studied crystals allowing to obtain clear diffraction spots for the freshly-cleaved surface of insulating NiPS$_3$. Further results obtained using electron spectroscopy methods indicate strong correlations effects in this material and we show that all electron spectroscopy data can be successfully described using cluster based approach for the Ni$^{2+}$ ion in octahedral field. NEXAFS Ni\,$L_{2,3}$ data show that NiPS$_3$ can be described as a transfer-charge insulator with $U_{dd}>\Delta$. This is also supported by the systematic $2p\rightarrow3d$ ResPES data collected around the Ni\,$L_{2,3}$ absorption edge, which clearly demonstrate the strong resonant behavior for the $3d^{n-1}$ photoemission final state. As deduced from these data, this final state is located at large binding energies compared to those for $3d^{n}\underline{L}$ final state located at the top of the valence band. The present results are first systematic electron spectroscopy data allowing to understand the electronic structure of these new layered materials and are of paramount importance for the further studies giving a strong basis for the understanding their transport, optical, and catalytic properties. 

\begin{acknowledgement}
This work was supported by the National Natural Science Foundation of China (Grant No. 21973059). A.M. acknowledges the BMBF (grant No. 05K19KER). D.S. acknowledges the BMBF (grant no. 0519ODR). A.T. acknowledges the Ministry of Science and Higher Education of the Russian Federation no. 0852-2020-0019 (State assignment in the field of scientific activity, Southern Federal University, 2020 project no. BAZ0110/20-1-03EH) and resources of the center of the collective use of SFedU ``High-Resolution Electron Microscopy''. We thank HZB for the allocation of synchrotron radiation beamtime and for support within the bilateral Russian-German Laboratory program. The North-German Supercomputing Alliance (HLRN) is acknowledged for providing computer time.
\end{acknowledgement}

\begin{suppinfo}
The following files are available free of charge
\begin{itemize}
  \item Description of experimental methods, sample preparation and DFT calculations. Additional experimental and theoretical data (PDF) can be downloaded via link: https://pubs.acs.org/doi/10.1021/acs.jpclett.1c00394
\end{itemize}

\end{suppinfo}

%\bibliography{/Users/YuDedkov/WORK/Articles/___REFERENCES___/references_all.bib}

\providecommand{\latin}[1]{#1}
\makeatletter
\providecommand{\doi}
  {\begingroup\let\do\@makeother\dospecials
  \catcode`\{=1 \catcode`\}=2 \doi@aux}
\providecommand{\doi@aux}[1]{\endgroup\texttt{#1}}
\makeatother
\providecommand*\mcitethebibliography{\thebibliography}
\csname @ifundefined\endcsname{endmcitethebibliography}
  {\let\endmcitethebibliography\endthebibliography}{}

%%%

\clearpage

\clearpage

\clearpage

% Table of contents entry should be 50 - 60 words long
% Image should be 55 mm broad and 50 mm high or 110 mm broad and 20 mm high

%\begin{figure}
%\textbf{Table of Contents}\\
%\medskip
%  \includegraphics{toc-image.png}
%  \medskip
%  \caption*{ToC Entry}
%\end{figure}

\end{document}